\pdfoutput=1
\documentclass[aps,prb,twocolumn,superscriptaddress,10pt]{revtex4-1}

\usepackage{amsmath}
\usepackage{graphicx}
\usepackage{bm}

\begin{document}


\title{Orbital magnetization in dilute ferromagnetic semiconductors}


\author{Cezary \'{S}liwa}
\affiliation{Institute of Physics, Polish Academy of Sciences,
aleja Lotnik\'{o}w 32/46, PL-02-668 Warszawa, Poland}

\author{Tomasz Dietl}
\affiliation{Institute of Physics, Polish Academy of Sciences,
aleja Lotnik\'{o}w 32/46, PL-02-668 Warszawa, Poland}
\affiliation{Institute of Theoretical Physics, University of Warsaw,
ulica Ho\.{z}a 69, PL-00-681 Warszawa, Poland}
\affiliation{WPI-Advanced Institute for Materials Research (WPI-AIMR),
Tohoku University, 2-1-1 Katahira, Aoba-ku, Sendai 980-8577, Japan}

\date{\today}

\begin{abstract}
The relationship between the modern and classical Landau's approach to carrier orbital magnetization is studied theoretically within the envelope function approximation, taking ferromagnetic (Ga,Mn)As as an example. It is shown that while the evaluation of hole magnetization within the modern theory does not require information on the band structure in a magnetic field, the number of basis wave functions must be much larger than in the Landau approach to achieve the same quantitative accuracy. A numerically efficient method is proposed, which takes advantages of these two theoretical schemes. The computed magnitude of orbital magnetization is in accord with experimental values obtained by x-ray magnetic circular dichroism in (III,Mn)V compounds. The direct effect of the magnetic field on the hole spectrum is studied too, and employed to interpret a dependence of the Coulomb blockade maxima on the magnetic field in a single electron transistor with a (Ga,Mn)As gate.
\end{abstract}

\pacs{}

\maketitle

\section{Introduction}

The last decade has witnessed the discovery of striking phenomena
associated with geometric and topological aspects of the band structure,
brought about by the
presence of spin-orbit coupling and the breaking of spin rotation symmetry.\cite{Bruno:2007_B,Resta:2010_JPCM}
In the case of ferromagnets the Berry curvature of bands hosting spin-polarized
carriers was found to result in sizable contributions
to transport coefficients, such as the anomalous Hall conductance.\cite{Jungwirth:2002_PRL,Nagaosa:2010_RMP}
It has been suggested more recently that the Berry curvature also describes
the orbital part of carrier magnetization coming from delocalized circulation.\cite{Xiao:2005_PRL,Thonhauser:2005_PRL,Shi:2007_PRL,Thonhauser:2011_IJMPB,Zhu:2012_PRB,Schulz-Baldes:2013_CMP}
This recent development is particularly worthwhile, as it has delivered formulas for
carrier orbital magnetization in the form that can be directly implemented into {\em ab initio} methods,
allowing us to interpret theoretically experimental values of the orbital magnetic moment provided by, for instance, x-ray circular magnetic dichroism (XMCD).\cite{Ceresoli:2010_PRB,Lopez:2012_PRB} The verification of the modern theory in this way is especially
meaningful since, compared to transport coefficients, thermodynamic properties are less sensitive
to scattering and localization.

In this paper, we examine quantitatively magnetization of spin-polarized valence band holes in
dilute ferromagnetic semiconductors (DFSs).\cite{Dietl:2014_RMP} A particular versatile method
to model semiconductor properties and devices is the Kohn-Luttinger (KL) envelope function approximation,\cite{Winkler:2003_B}
whose six-band version has been exploited to describe various thermodynamic\cite{Dietl:2014_RMP}
and transport data in DFSs, including the anomalous Hall effect.\cite{Jungwirth:2002_PRL,Nagaosa:2010_RMP} Within this scheme, we
compare carrier magnetization obtained from the modern approach\cite{Xiao:2005_PRL,Thonhauser:2005_PRL,Shi:2007_PRL,Thonhauser:2011_IJMPB,Zhu:2012_PRB,Schulz-Baldes:2013_CMP}
and determined\cite{Dietl:2001_PRB,Jungwirth:2006_PRB,Sliwa:2006_PRB} employing the time-honored Landau theory.\cite{Landau:1930_ZP}

According to the combined KL and Landau's method, the spin-orbit interaction generates two contributions to orbital magnetization $M_{\text{orb}}$ in DFSs.\cite{Dietl:2001_PRB}
The first one, $M_{\text{L}}$, stems from Landau's quantization. The second contribution $M_{I}$ is proportional to the orbital angular momentum operator $\mathbf{\hat I}$.
As we demonstrate here, only $M_{\text{L}}$ is reproduced by the modern approach, but the second term emerges within the modern approach if the set of the basis of the Bloch wave functions $\{u_n\}$ is enlarged.
Furthermore, we show that both contributions have to be taken into account to describe quantitatively
experimental results
on XMCD,\cite{Freeman:2008_PRB,Wadley:2010_PRB} and on
the dependence of the chemical potential on the magnetic field  in (Ga,Mn)As.\cite{Ciccarelli:2012_APL}

\section{Landau theory within the envelope function approach}

Within the KL method and neglecting the lack of inversion symmetry, the six-band Hamiltonian of holes in a magnetic field $\mathbf{B}$ and in the presence of Mn magnetization $\mathbf{M}$ consists of three terms in DFSs:\cite{Dietl:2001_PRB} (i) $\mathcal{{H}}_{\text{L}}$ that describes Landau's
quantization of the valence band in terms of the Luttinger band structure parameters $\gamma_1$, $\gamma_2$, and $\gamma_3$; (ii) the Zeeman-like contribution $H_{\text{Z}}$, and (iii) $\mathcal{H}_{pd}$ accounting for $p$-$d$ coupling between hole and Mn spins. In the basis employed previously,\cite{Winkler:2003_B,Sliwa:2006_PRB,Trebin:1979_PRB}
\begin{equation}
  \mathcal{{H}}_{\text{Z}} = - (1 + 3\kappa)\mu_B \mathbf{\hat I}\cdot \mathbf{B}
	  + g_0 \mu_B \mathbf{\hat s}\cdot \mathbf{B},
\label{eq:H_Z}
\end{equation}
where $\kappa$ is one more Luttinger parameter,\cite{Luttinger:1956_PR,Roth:1959_PR,Winkler:2003_B} the free electron Land\'e factor $g_0 \approx 2.002$, and the dimensionless angular-momentum tensor operators $\mathbf{\hat I}$ and~$\mathbf{\hat s}$ are given by,
\begin{equation}
  \mathbf{\hat I} = \begin{pmatrix} \displaystyle \frac{2}{3} \mathbf{\hat J}&
	    \mathbf{\hat U} \\ \mathbf{\hat T}&
			\displaystyle \frac{2}{3} \mathbf{\hat{\boldsymbol{\sigma}}}
		\end{pmatrix}, \qquad
  \mathbf{\hat s} = \begin{pmatrix} \displaystyle \frac{1}{3} \mathbf{\hat J}&
	    -\mathbf{\hat U} \\ -\mathbf{\hat T}&
			\displaystyle -\frac{1}{6} \mathbf{\hat{\bm{\sigma}}}
		\end{pmatrix}.
\end{equation}
Following Ref.~\onlinecite{Trebin:1979_PRB}, we denote by $\mathbf{\hat{\bm{\sigma}}}$
the Pauli matrices, by $\mathbf{\hat J}$ the set of spin-$3/2$ angular-momentum matrices,
and by $\mathbf{\hat U}$, $\mathbf{\hat T}$ the sets of matrices for the cross-space.
In Eq.\,(\ref{eq:H_Z}), besides the ordinary Pauli spin part $g_0 \mu_B \mathbf{\hat s}\cdot \mathbf{B}$, there is an orbital term
$\mathcal{{H}}_{I} = -(1 + 3\kappa)\mu_B\mathbf{\hat I}\cdot\mathbf{B}$. This contribution is brought about by coupling of the six valence subbands to remote bands in the presence of an external magnetic field.
That is,
$H_I$ accounts for an admixture of the orbital magnetic moment to the carrier effective Land\'e factor.\cite{Roth:1959_PR}

Finally, the $p$-$d$ coupling to the spin-polarized Mn ions is taken into account in the virtual-crystal and
molecular-field approximations, leading to additional giant spin splitting of Landau levels, described
by the Hamiltonian $\mathcal{H}_{pd} = (\Delta_{\text{v}}/ M )\mathbf{M} \cdot \mathbf{\hat s}$,
where $\Delta_{\text{v}}$ is the $p$-$d$ exchange splitting
of the valence band top.\cite{Jungwirth:2002_PRL,Dietl:2001_PRB,Sliwa:2006_PRB}

Within Landau's method\cite{Landau:1930_ZP} the carrier
magnetization $\mathbf{M}_{\text{c}}(T,\mathbf{B})$ is given by
the derivative of the grand thermodynamic potential,
\begin{eqnarray}
\label{eq:M_Landau}
  	\lefteqn{\Omega_{\text{c}} = -\mu_{\text{B}} B k_{\text{B}}T
     \sum_j\int\limits_{-\infty}^{\infty} \frac{m_0\, dk_3}{2(\pi\hbar)^2}
	  } \\ & & \qquad \qquad 
     \ln\left\{1+\exp \left( -[E_j(k_3) - \mu]/k_{\text{B}}T \right)\right\},
     \nonumber
\end{eqnarray}
with respect to the magnetic field,
$\mathbf{M}_{\text{c}} = -\partial \Omega_{\text{c}}/\partial \mathbf{B}$, where $m_0$ in Eq.\,(\ref{eq:M_Landau}) is the free-electron mass.
Here,
$E_j(k_3)$ is the $j$-th eigenenergy of $\mathcal{H}_{\text{L}}+\mathcal{H}_{\text{Z}} + \mathcal{H}_{pd}$
for a carrier
with the $\mathbf{k}$ component along the direction of the
magnetic field denoted as $k_3$, and $\mu$ is the chemical potential. The values of $\mathbf{M}_{\text{c}}$
computed in this way for (Ga,Mn)As were reported previously.\cite{Sliwa:2006_PRB}

This approach allows us to evaluate orbital parts of carrier magnetization, $M_{\text{L}}$ and $M_{I}$, associated with $\mathcal{H}_{\text{L}}$ and $\mathcal{H}_{\text{I}}$, respectively, at a given $p$-$d$ exchange splitting of bands described by $\mathcal{H}_{pd}$. The key question we address in this paper is how these two contributions are related to orbital magnetization $M_{\text{mod}}$ obtained from the modern theory. A formulation of the modern theory within the KL method is discussed in the subsequent section.

\section{Modern theory of orbital magnetization}

Within the modern approach the orbital part of $\mathbf{M}_{\text{c}}$ at $B = 0$ for $N$ bands is given by,
\begin{eqnarray}
  \label{eq: modern}
  \lefteqn{\mathbf{M}_{\text{mod}} = \mu_{\text{B}}
	   \int \frac{d^3\mathbf{k}}{(2\pi)^3} \sum_{n, n' = 1}^N \mathcal{M}_T(E_{n'\mathbf{k}}, E_{n\mathbf{k}}) } \\ 
		& & \qquad \qquad
		   \mathop{\mathrm{Im} } \left[ 
		  m_0
      \left\langle u_{n\mathbf{k}} \middle| \mathbf{\hat{v}} \middle| u_{n'\mathbf{k}} \right\rangle
      \times
      \left\langle u_{n'\mathbf{k}} \middle| \mathbf{\hat{v}} \middle| u_{n\mathbf{k}} \right\rangle \right],
	  \nonumber
\end{eqnarray}
where $u_{n\mathbf{k}}$ is 
the Bloch function corresponding to the eigenenergy $E_{n\mathbf{k}}$
of the KL Hamiltonian $\mathcal{H}_{\mathbf{k}}$ at $B =0$ including $\mathcal{{H}}_{pd}$;
$\hbar \mathbf{v} = \partial \mathcal{{H}}_{\mathbf{k}}/\partial \mathbf{k}$,
and
\begin{eqnarray}
  \lefteqn{\mathcal{M}_T(E_{n'\mathbf{k}}, E_{n\mathbf{k}}) = {}} \nonumber \\ & & \qquad
    \mathcal{M}\bigl[ (E_{n'\mathbf{k}}-\mu)/k_{\text{B}}T,
		  (E_{n\mathbf{k}}-\mu)/k_{\text{B}}T \bigr] / k_{\text{B}}T,
\end{eqnarray}
where the dimensionless function~$\mathcal{M}$ reads
\begin{eqnarray}
\label{eq:m}
  \lefteqn{\mathcal{M}(x_{n'}, x_n) = \frac{1}{x_{n'} - x_n} \biggl[
    \frac{f(x_{n'}) + f(x_n)}{2} + {}} \\ & & \qquad {} + \frac{
      \ln[1+\exp(-x_{n'})] - \ln[1+\exp(-x_n)]}{x_{n'} - x_n} \biggr], \nonumber
\end{eqnarray}
with the Fermi-Dirac distribution function $f(x) =[1 + \exp(x)]^{-1}$
(notice that the cross product of velocity matrix elements is purely imaginary).

\begin{figure}[tb]
  \includegraphics[width=3in]{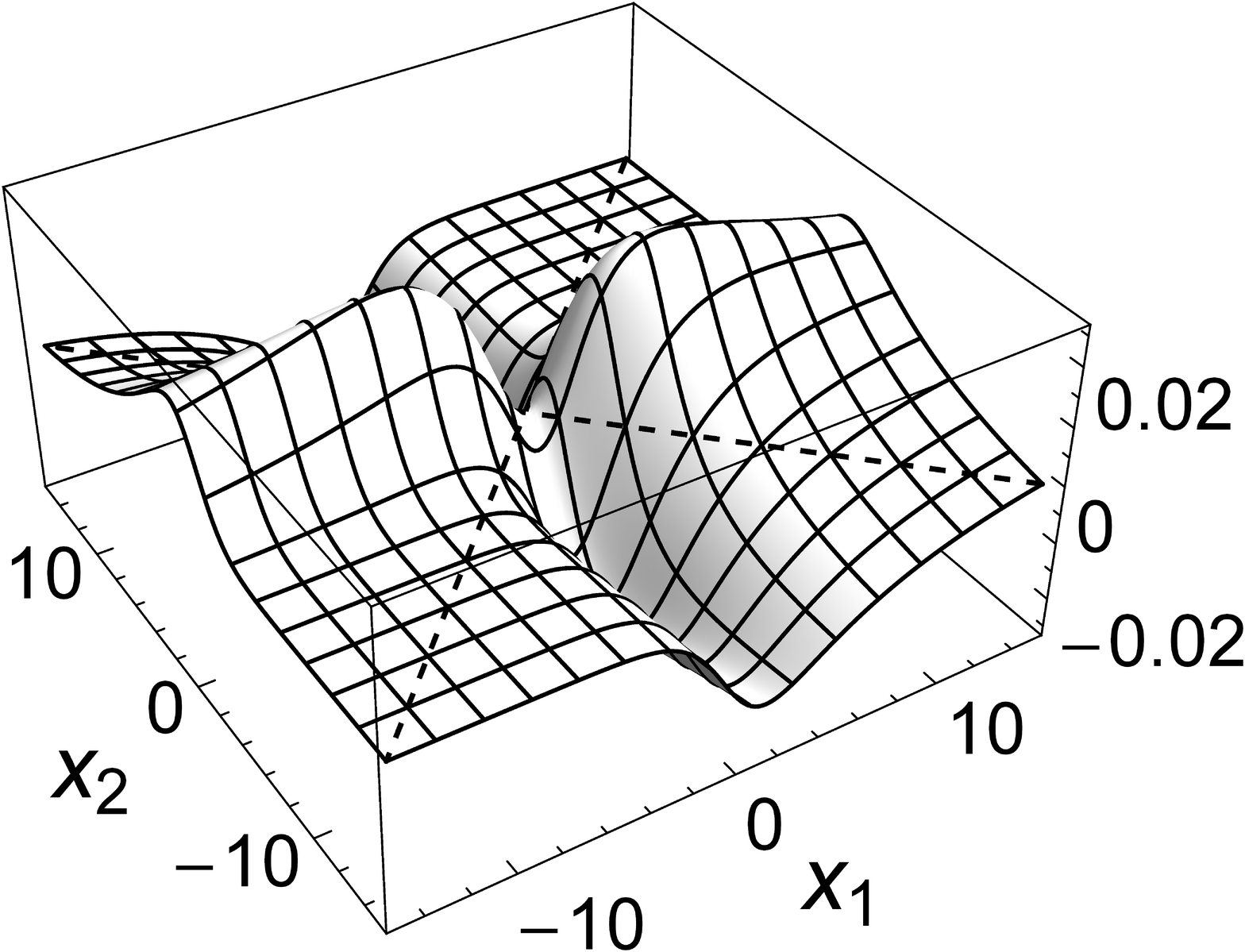}
	\caption{Plot of the function $\mathcal{M}(x_1,x_2)$  that appears in the formula [Eq.\,(\ref{eq:m})] for orbital magnetization.}
  \label{fig: m}
\end{figure}

The contribution coming from the first term in Eq.\,(\ref{eq:m}) corresponds to magnetization of the carriers' wave packets,\cite{Chang:2008_JPCM} whereas the second is proportional to the Berry curvature.
The definition of $x$ implies that $x =0$ for states at the Fermi level, whereas $x > 0$ and $x < 0$ correspond to the empty and occupied states, respectively.
Since the formula for magnetization involves a symmetric summation over a pair of indices running over the same set of bands, and the cross product is antisymmetric, we have adopted ${\mathcal{M}}(x_1, x_2)$ in an antisymmetrized form that allows us to tackle better with a possible divergence at $x_1 = x_2$.  As seen in Fig.~\ref{fig: m},  $\mathcal{M}(x_2, x_1) = - \mathcal{M}(x_1, x_2)$,
and $\mathcal{M}$ vanishes rather than diverges at the band crossings, $\mathcal{M}(x, x) =0$, as required for degenerate bands.
The function $\mathcal{M}(x_1, x_2)$
also obeys $\mathcal{M}(-x_1, -x_2) = \mathcal{M}(x_1, x_2)$ (electron-hole symmetry) and $\mathcal{M}(x, -x) = 0$.
Furthermore, according to Fig.~\ref{fig: m}, $\mathcal{M}(x_1, x_2)$ decays exponentially to zero with inverse temperature in the first and third quadrant, i.e., when $x_1$ and $x_2$ have the same sign (either
positive or negative, corresponding to pairs of empty or pairs of occupied states, respectively).  This formulation substantiates
a picture in which orbital magnetization is described by a sum
over \emph{pairs} of subbands, with significant contributions only from empty--occupied states.

\section{Comparison of the two approaches}

\begin{figure*}
  \leavevmode
    \hbox to 0.2\hsize{\hspace{1.25em}(a)\hfill}\hfill
	  \hbox to 0.2\hsize{\hspace{1.25em}(b)\hfill}\hfill
    \hbox to 0.2\hsize{\hspace{1.25em}(c)\hfill}\hfill
    \hbox to 0pt{}\\[0.2em]
  \hbox to \hsize{%
    \hfill\includegraphics[scale=0.7]{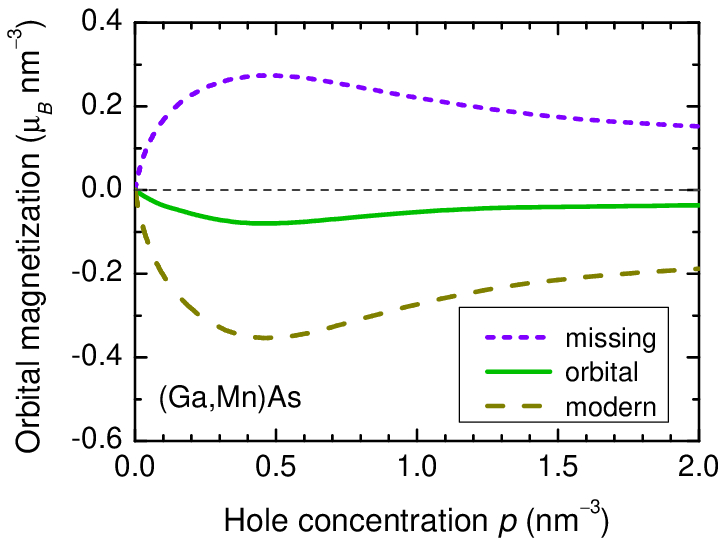}\hfill
    \hfill\includegraphics[scale=0.7]{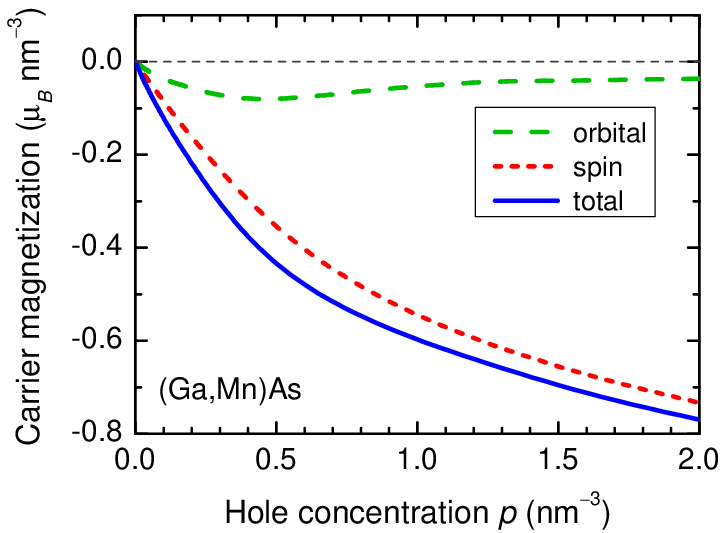}\hfill
    \hfill\includegraphics[scale=0.7]{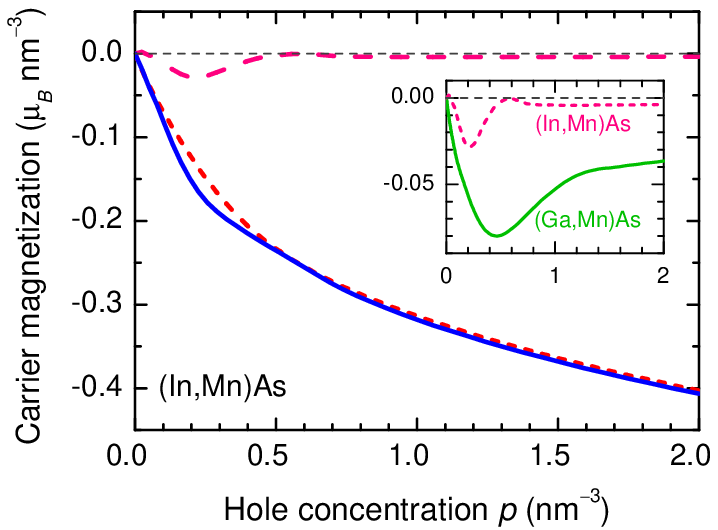}\hfill
	}
  \caption{(Color online) Contributions to hole magnetization at 10\,K computed within the six-band Kohn-Luttinger model of the valence band for parameters of GaAs ($\gamma_1 = 6.85$, $\gamma_2 = 2.1$, $\gamma_3 = 2.9$, $\kappa = 1.2$, $\Delta_{\text{SO}} = 0.341 \, \mathrm{eV}$) and InAs ($\gamma_1 = 20.0$, $\gamma_2 = 8.5$, $\gamma_3 = 9.2$, $\Delta_{SO} = 0.39 \, \mathrm{eV}$, $\kappa = 7.60$), and $\mathbf{M}\parallel \langle 100\rangle$ and the parameter of valence band exchange splitting $\Delta_{\text{v}} = -180 \, \mathrm{meV}$, corresponding to the magnitude of saturation magnetization in $\mathrm{Ga}_{0.95}\mathrm{Mn}_{0.05}\mathrm{As}$. Orbital magnetization $M_{\text{orb}}$ in (Ga,Mn)As from Landau's method (solid line) is decomposed into $M_{\text{L}}= M_{\text{mod}}$ provided by the modern method and the remaining (missing) part $M_{I}$ (dashed and dotted lines, respectively) (a).  Total hole magnetization $M_{\text{c}}$ (solid line) in (Ga,Mn)As (b) and (In,Mn)As (c) decomposed into $M_{\text{orb}}$  and the spin part $M_{\text{spin}}$ (dashed and dotted lines, respectively). Inset in (c) shows $M_{\text{orb}}$ for (In,Mn)As and (Ga,Mn)As in an expanded scale.}
    \label{fig:six_bands}
\end{figure*}

We first compare hole orbital magnetization determined within the KL method from the modern approach, $M_{\text{mod}}$  [Eq.\,(\ref{eq: modern})] to $M_{\text{L}}$ determined from the grand thermodynamic potential [Eq.\,(\ref{eq:M_Landau})] in the limit $B \to 0$. In order to evaluate $M_{\text{L}}$, i.e., orbital magnetization resulting from Landau quantization of the hole spectrum, we assume $\mathcal{{H}}_{Z} =0$, i.e., neglect the contribution $M_I$ to orbital magnetization. For $\mathrm{Ga}_{1-x}\mathrm{Mn}_{x}\mathrm{As}$, in the explored parameter space ($T = 10 \, \mathrm{K}$, $5\times10^{19} \leq p \leq 10^{21} \, \mathrm{cm}^{-3}$, and $\Delta_{\text{v}} =-180\,\mathrm{meV}$, i.e., $x \simeq 0.05$), the relative difference between the data obtained by these two methods is within our numerical uncertainty of $10^{-5}$. This finding highlights a major progress provided by the modern approach that allows one to circumvent the computational load associated with the determination of Landau level energies for complex band structures.

However, quantitative agreement between the Landau and modern approach to orbital magnetization is obtained neglecting $M_{I}$. This indicates that the term arising from the coupling to remote bands, $-(1 + 3\kappa)\mu_B\mathbf{\hat I}$, is not taken into account within the modern approach. The magnitude of the missing magnetization $M_{I}$ can be evaluated from the grand thermodynamic potential [Eq.\,(\ref{eq:M_Landau})] with eigenenergies $E_{n\mathbf{k}}$ of the Hamiltonian ${\mathcal{H}}_{\mathbf{k}} + {\mathcal{H}}_{I}$. According to results presented in Fig.\,\ref{fig:six_bands}(a), $M_{I}$ is quite sizable and, in fact, compensates largely $M_{\text{L}} = M_{\text{mod}}$ provided by the modern approach.  For comparison, we also show the total hole magnetization $M_{\text{c}}$ that is seen to be dominated by the spin part $M_{\text{spin}}$, obtained from ${\mathcal{H}}_{\mathbf{k}} + g_0\mu_B\mathbf{\hat s}\cdot{\mathbf{B}}$, in both (Ga,Mn)As [Fig.\,\ref{fig:six_bands}(b)] and (In,Mn)As [Fig.\,\ref{fig:six_bands}(c)].

\section{Discussion}

The results presented in the previous section point to disagreement between the two theories of orbital magnetization. A question then arises whether $M_{I}$ is an artifact of the Landau approach or rather it is the modern theory that disregards the quantitatively important contribution $M_{I}$.

In order to address this issue we note that the modern approach requires information on both eigenenergies and eigenfunctions. In contrast, the Landau method is developed in terms of eigenenergies only. Within the KL method, the second order perturbation theory serves to determine the contribution to carrier eigenenergies of bands beyond the valence band states. However, no effect of the remote bands on the eigenfunctions $u_{n\mathbf{k}}$ is considered within such an approach. This suggests that by taking into account a contribution of remote bands to $u_{n\mathbf{k}}$, either perturbatively or by enlarging the basis $\{u_n\}$ of the KL scheme, the accuracy of the modern approach can be improved. To verify this hypothesis we have computed the magnitude of orbital magnetization within the eight-band model that incorporates the conduction band states to $\{u_n\}$.\cite{Winkler:2003_B}

Within the six-band model the grand thermodynamic potential has been derived in the hole picture. Since in the eight-band model the energies are bound neither from below nor from above, it is necessary to use the electron picture in order to describe the states residing above a fixed energy in the band gap. That is, we exploit the identity $-\ln(1+e^{-x}) = -\ln(1+e^{x}) + x$ to split the grand thermodynamic potential into a sum of the hole contribution and a hole-concentration independent shift. The shift describes the magnitude of orbital magnetization for the fully occupied valence band brought about by transitions to the conduction band, and it vanishes in the absence of band spin splittings.

Within the modern approach, an equivalent approach is to decompose $\mathcal{M}(x_{n'}, x_n)$ as follows:
\begin{eqnarray}
  \lefteqn{\mathcal{M}(x_{n'}, x_n) = \frac{1}{x_{n'} - x_n} \biggl[
    \frac{f(x_{n'}) - f(-x_n)}{2} + {}} \\ & & \qquad {} + \frac{
      \ln(1+e^{-x_{n'}}) - \ln(1+e^{x_n})}{x_{n'} - x_n}
			\biggr] + \frac{1}{2} \frac{x_{n'} + x_n}{(x_{n'} - x_n)^2} . \nonumber
\end{eqnarray}
As can be shown by inspection, also here the second term leads to a shift independent of the hole concentration but dependent on band spin splittings; it assumes a nonzero value if spin splittings of the valence and conduction bands differ, $\Delta_{\text{v}} \ne \Delta_{\text{c}}$. Since it provides just an additional contribution to the magnitude of orbital magnetization coming from fully occupied bands, a comparison between the two approaches is still meaningful even if we disregard the shift.

\begin{figure}[b]
  \includegraphics[scale=1.0]{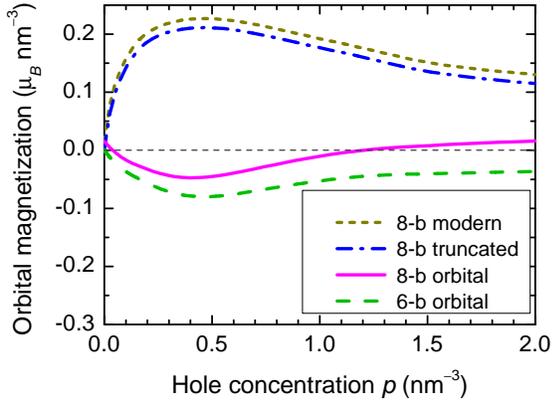}
  \caption{(Color online) Orbital magnetization $M_{\text{orb}}$ of (Ga,Mn)As computed
	by the Landau method within the eight- and six-band models (solid and dashed lines,
    respectively). These results differ substantially from the outcome
    of the eight-band modern model (the dotted line).
    However, the modern model (except for a concentration-independent shift describing magnetization of the
    fully occupied band) agrees with the truncated eight-band Landau model
    (no coupling to remote bands; dash-dotted line).
    The computations have been carried out for splitting of the valence and conduction
    bands, $\Delta_{\text{v}} = -180$\,meV and $\Delta_{\text{c}} = 30$\,meV,
    respectively.}
  \label{fig:eight_bands}
\end{figure}

As shown in Fig.\,\ref{fig:eight_bands}, within the Landau theory there is a minor change in the magnitudes of $M_{\text{orb}}$ on going from the six- to the eight-band model, as eigenenergies are fairly accurately provided by either of these two KL schemes. In contrast, there is a considerable difference between magnetization values for these two KL implementations within the modern approach, as seen comparing the data in Figs.\,\ref{fig:six_bands} and \ref{fig:eight_bands}. This demonstrates that the enlargement of the set $\{u_n\}$ has a substantial influence on the magnitude of $M_{\text{mod}}$. However, according to the data in Fig.\,\ref{fig:eight_bands}, $M_{\text{mod}}$ obtained in this way still disagrees with $M_{\text{orb}}$ from the Landau method. Actually, according to the results in Fig.\,\ref{fig:eight_bands}, the modern method is in accord with a truncated variant of the eight-band Landau method, in which the coupling to bands beyond the eight-band manifold is disregarded (i.e. $\kappa\prime = -1/3$ in the notation of Ref.\,\onlinecite{Winkler:2003_B}). This indicates that for the modern approach the eight-band basis is still too small for obtaining accurate values of orbital magnetization.

Altogether these findings imply that it is possible to determine orbital magnetization
without referring to carrier spectrum in the magnetic field but to achieve the
same quantitative accuracy the set of basis wave functions $\{u_{n}\}$
must be much larger in the modern approach than needed within the Landau theory.
 However, there exists an efficient  method
to compute $M_{\text{orb}}$ at $B = 0$ exploiting advantages of
these two theoretical schemes. The hybrid method we propose consists of evaluating
orbital magnetization as
$M_{\text{orb}}  = M_{\text{mod}} + M_I$,
where both $M_{\text{mod}}$ and  $M_I$ are to be computed within the minimal KL scheme
for the problem at hand (typically either six- or eight-band model).
Thus, the hybrid method requires only a small set of basis wave functions $\{u_{n}\}$ and supplies accurate values of $M_{\text{orb}}$ without
computing Landau level energies. Below, we compare experimental data for (Ga,Mn)As to our theoretical results obtained by the hybrid procedure within the eight-band KL scheme.

\section{Comparison to available experimental data}

\begin{figure}[b]
  \includegraphics[scale=0.8]{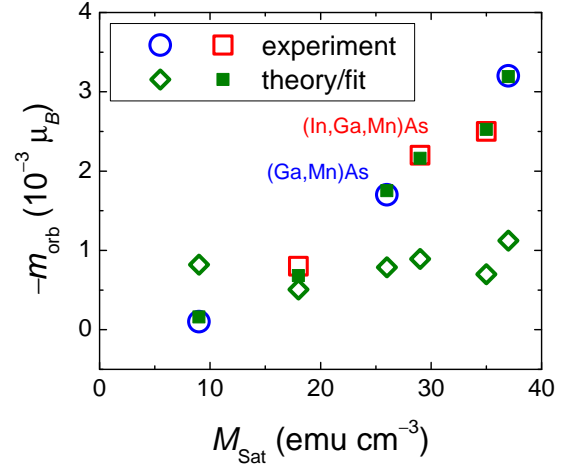}
	\caption{(Color online) Orbital magnetization of As $4p$ states
determined experimentally at $\sim 10$\,K and in 2\,T by Wadley {\em et
al.}\cite{Wadley:2010_PRB} as a function of saturation magnetization
$M_{\text{Sat}}$ for (Ga,Mn)As (open circles) and (In,Ga,Mn)As (open squares)
compared to theoretical values of orbital magnetization $M_{\text{orb}}$
obtained from the hybrid method within the eight-band KL model for (Ga,Mn)As directly (empty
diamonds) and including a possible contribution $\Delta m_{\text{orb}}
= a \Delta_{\text{v}} + b$, where $a$ and $b$ are fitting parameters
(full squares).}
	\label{fig:comparison}
\end{figure}

Figure \ref{fig:comparison} presents the orbital moment of As $4p$ states
determined by XMCD for (Ga,Mn)As and (In,Ga,Mn)As films with different
saturation magnetizations $M_{\text{Sat}}$ and Curie temperatures
$T_{\text{C}}$.\cite{Wadley:2010_PRB} Since the orbital moment of cations
appears to be much smaller,\cite{Wadley:2010_PRB,Freeman:2008_PRB} we
compare these data to our theory, evaluating $\Delta_{\text{v}}$ and hole
concentrations $p$ from $M_{\text{Sat}}$ and $T_{\text{C}}$ within the
eight-band $sp$-$d$ Zener model.\cite{Hankiewicz:2004_PRB,Winkler:2003_B} As seen, our theory explains
both the sign and the small magnitude of $m_{\text{orb}} =
M_{\text{orb}}/N_0$ observed experimentally, where $N_0$ is the anion
concentration. Since contributions to $m_{\text{orb}}$ coming from the
cations and fully occupied bands have been neglected, we may
expect an additional term proportional to $\Delta_{\text{v}}$.
Furthermore, experimental data were taken in 2\,T. This may lead to a
diamagnetic shift of $M_{\text{orb}}$, which should weakly depend on
$\Delta_{\text{v}}$. Accordingly, we supplement the theoretical values
of $m_{\text{orb}}$ with $\Delta m_{\text{orb}} = a\Delta_{\text{v}} + b$.
The fitting procedure implies $a = -14.8\times 10^{-3}$\,$\mu_{\text{B}}$/eV and $b
= -1.59 \times 10^{-3}$\,$\mu_{\text{B}}$. Although the quality
of the fit is excellent, such a large value of the offset~$b$
calls for further attention.

\begin{figure}[tb]
  \includegraphics[width=7cm]{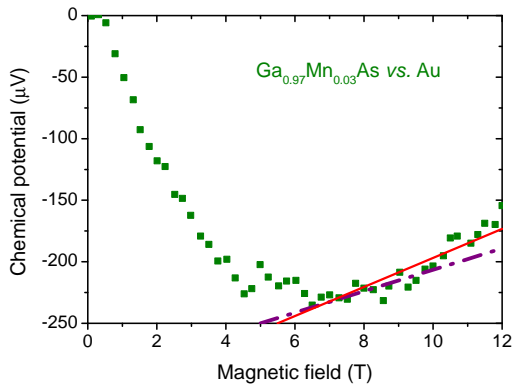}
	\caption{(Color online) Chemical potential determined experimentally for $\mathrm{Ga}_{0.97}\mathrm{Mn}_{0.03}\mathrm{As}$ at 0.3 K (squares) by Ciccarelli {\em et al}.\cite{Ciccarelli:2012_APL} Slopes of dashed and solid lines are computed for interstitial concentrations
$x_{\text{I}} = 0$ and 0.5\%, respectively.}	\label{fig:SET}
\end{figure}

Another relevant experiment concerns variations of the chemical potential
$\mu$ with the magnetic field $B$, as provided by studies of
an Al single electron transistor (SET) with a
(Ga,Mn)As gate.\cite{Ciccarelli:2012_APL}
Figure \ref{fig:SET} shows $\mu(B)$ determined from the field-induced shift of Coulomb blockade
peaks for a SET with the $\mathrm{Ga}_{0.97}\mathrm{Mn}_{0.03}\mathrm{As}$ gate
in respect to the shift in a control SET with an Au gate.\cite{Ciccarelli:2012_APL}
We are interested in the region $B \gtrsim 7$\,T, in which the Mn spins become saturated but
nevertheless $\mu$  varies with the magnetic field.

In order to explain these data we make use of relations $\mu = \partial \Omega_{\text{c}}/\partial p$
and $M_c = - \partial \Omega_{\text{c}}/\partial B$, which lead to the thermodynamic identity,
\begin{equation}
  \left.\frac{\partial \mu}{\partial B}\right|_p =
          \frac{\partial^2 \Omega_\text{c}}{\partial B \, \partial p} =
          -\left.\frac{\partial M_{\text{c}}}{\partial p}\right|_B,
        \label{first}
\end{equation}
implying $\partial M_{\text{c}} / \partial p = -\partial \mu / \partial B$,
which relates the derivative of the carrier magnetization (with respect to carrier
concentration) to changes of the electron's chemical potential in an external
magnetic field.

Because of virtual cancellations between $M_{\text{mod}}$
and $M_{I}$, the total hole magnetization $M_{\text{c}}$ is dominated
by the spin part [see, Fig.~\ref{fig:six_bands}(b)] that is isotropic.
This explains why $\mu(B)$ was independent of the field direction in
respect to crystallographic axes.\cite{Ciccarelli:2012_APL} In order
to evaluate $\partial M_{\text{c}} / \partial p$ information on
saturation magnetization and hole concentration are needed, which at given $x$
depend on density of Mn interstitials $x_{\text{I}}$.\cite{Dietl:2014_RMP}
As shown in Fig.\,\ref{fig:SET}, theoretical results obtained for
$x_{\text{I}} = 0$ and 0.5\% are consistent with the experimental data.

\section{Conclusions}

In summary, we have proposed a numerically efficient method that
combines advantages of the modern and Landau approach to carrier orbital magnetization. The computed hole magnetization within
the formalism developed here explains the magnitude of orbital and spin magnetizations implied
by experimental studies of XMCD and the Coulomb blockade in (Ga,Mn)As.
A timely question arises about implications of our findings to the theory
of anomalous and spin Hall effects in semiconductors.

\section*{Acknowledgments}
We thank B.\,L. Gallagher, K.\,Edmonds, and P.\,Wadley for instructive discussions on XMCD results.
This work was supported by the European Research Council through the FunDMS Advanced
Grant (No. 227690) within the ``Ideas'' 7th Framework Programme of the EC.

\bibliography{magcorb}

\end{document}